\documentclass{sigchi-ext}
\usepackage[T1]{fontenc}
\usepackage{textcomp}
\usepackage[scaled=.92]{helvet} 
\usepackage{graphicx} 
\usepackage{balance}  
\usepackage{booktabs} 
\usepackage{ccicons}  
\usepackage{ragged2e} 

\def\plaintitle{SIGCHI Extended Abstracts Sample File: Note Initial
  Caps} 
\def\emptyauthor{}
\def\plainkeywords{privacy; ethics; mixed reality; eye tracking}

\title{Privacy Implications of Eye Tracking in Mixed Reality}

\numberofauthors{6}
\author{%
  \alignauthor{%
    \textbf{Diane Hosfelt}\\
    \affaddr{Mozilla} \\
    \email{dhosfelt@mozilla.com} }\alignauthor{%
    \textbf{Nicole Shadowen}\\
    \affaddr{Mozilla}\\
    \email{nshadowen@mozilla.com} } \vfil \alignauthor{%
    } }

\definecolor{linkColor}{RGB}{6,125,233}
\hypersetup{%
  pdftitle={\plaintitle},
  pdfauthor={\emptyauthor},
  pdfkeywords={\plainkeywords},
  bookmarksnumbered,
  pdfstartview={FitH},
  colorlinks,
  citecolor=black,
  filecolor=black,
  linkcolor=black,
  urlcolor=linkColor,
  breaklinks=true,
}

\begin{document}

\CopyrightYear{2020}
\setcopyright{rightsretained}
\conferenceinfo{CHI'20,}{April  25--30, 2020, Honolulu, HI, USA}
\isbn{978-1-4503-6819-3/20/04}
\doi{https://doi.org/10.1145/3334480.XXXXXXX}
\copyrightinfo{\acmcopyright}

\maketitle

\RaggedRight{} 

\begin{abstract}
  Mixed Reality (MR) devices require a world with always-on sensors and real-time processing applied to their outputs. We have grappled with some of the ethical concerns presented by this scenario, such as bystander privacy issues with smartphones and cameras. However, MR technologies demand that we define and defend privacy in this new paradigm. This paper focuses on the challenges presented by eye tracking and gaze tracking, techniques that have commonly been deployed in the HCI community for years but are now being integrated into MR devices by default.
\end{abstract}

\keywords{\plainkeywords}

\ccsdesc[500]{Human-centered computing~Mixed/Augmented Reality}
\ccsdesc[300]{Human-centered computing~Human computer interaction (HCI)}
\ccsdesc[100]{Social and professional topics~Code of ethics}

\printccsdesc

\section{Introduction}

Mixed reality devices blend digital elements and the physical world, covering a wide spectrum that includes both virtual reality (VR) and augmented reality (AR). In VR, a device occludes users' vision, and often other senses, to present a fully digital experience. On the other hand, AR experiences overlay digital elements on users' perceptions of the physical world~\cite{hosfelt}. This paper will focus on immersive MR, where a user's field of vision is entirely controlled by a head-mounted display (HMD). These devices have been in development for years, but the development of ethical frameworks for this field has only recently emerged. This work fits into the ``biometric data and identity'' category of the MR ethics framework presented by Bye (2019)~\cite{Bye}. We examines the risks of uniquely identifying users using interpupillary distance (IPD) and present privacy threats posed by eye tracking.

IPD is the distance between the center of the pupils of the eyes, and it is used to determine the positioning of the lenses in glasses. Improper positioning can cause eye fatigue and headaches. Just like in glasses or binoculars, each eye looks through a different lens in an HMD, making proper IPD configuration important, especially for long-term use. Devices such as the Oculus Quest have an adjustable IPD slider. Nevertheless, a camera pointed at the eyes, perhaps intended for gaze tracking, could also determine this metric.

In this paper, we differentiate between gaze tracking and eye tracking by noting that in \emph{gaze tracking}, the object of the gaze matters, whereas in \emph{eye tracking}, the actions of the eyes themselves matters. Eye tracking encompasses gaze tracking.

Eye tracking is a technique where a device measures an individual's eye movements, noting where a person is looking at any given time, as well as the order in which they shift their gaze from location to location. Our eyes reveal a map of our cognitive processes as we examine visual information, pausing on some, while skimming over other data. The HCI community has found eye tracking studies to be useful for providing insight into measuring situational awareness in air traffic control training~\cite{hauland2003measuring}, evaluating cockpit control design~\cite{hanson2004focus}, improving doctors' performance~\cite{law2004eye}, and determining if internet users look at banner ads on websites~\cite{albert2002web}.

\section{User identification}
Cross-site identification and user tracking has emerged as a serious problem in today's web. It violates users' agency and ability to choose what and how to share information about themselves with websites. IPD measurements are a distinguishable metric, particularly when combined with other information, such as IP addresses or device IDs, enabling simple user fingerprinting.

\begin{table}
\begin{tabular}{l l l}
     & Mean (mm) & Standard Deviation (mm)  \\
    Men & 64.0 & 3.4 \\
    Women & 61.7 & 3.6
\end{tabular}
\caption{IPD varies between individuals, and also varies for individuals at different focal lengths, offering numerous metrics for identification purposes}
\label{table:ipd}
\end{table}


Some modern MR devices, including the Oculus Quest, include an adjustable IPD slider for improved comfort, but any MR device that can track gaze can also compute IPD (see \autoref{table:ipd}~\cite{dodgson2004variation}). Pupil dilation may also be used to identify users~\cite{bednarik2005eye}.

Gaze can also uniquely identify individuals. The first Eye Movement Verification and Identification Competition (EMVIC) took place in 2012~\cite{kasprowski2012first}, using a jumping dot as the gaze target for participants. The competition provided four different datasets, differing in calibration and testing/training splits. The best models achieved accuracy ranging from 58\% to nearly 98\%. The second EMVIC took place in 2014 \cite{kasprowski2014second}. Each subject looked at 10 faces freely for 4 seconds, and were then identified using the entire session's information. The best models achieved nearly 40\% accuracy compared to a 3\% random guess probability. Further research is required, in particular to test identification using eye trackers that are currently shipping in MR headsets.

\section{Profiling and revealing sensitive characteristics}
The eyes make subconscious movements, betraying our inner thoughts. As its use in psychological research implies, gaze tracking can reveal sensitive characteristics about us, including our sexual orientation~\cite{bolmont2014love}. Research also indicates that gaze tracking can help diagnose disorders such as anxiety and depression~\cite{armstrong2012eye} and autism~\cite{boraston2007application}. This implies that an MR device with eye tracking could be misused to profile users against protected or sensitive characteristics, such as protected health data.

Consider browsing the immersive web and being targeted by advertisers who know that you are anxious. Perhaps they would use your anxiety against you to serve targeted ads that also claim that the deals are running out, knowing that people with your characteristics tend to be more susceptible to such techniques.

Our eyes can also indicate details of our decision making process. In particular, pupil dilation is linked to arousal---heightened arousal, such as that experienced during uncertainty, causes pupil dilation~\cite{murphy2014pupil}. Eye tracking can therefore enable uncertainty tracking. This could be used by advertisers to determine when you are wavering in your decision making process, allowing them to ``sweeten the deal'' and close the sale. In essence, uncertainty tracking enables additional manipulations.

The sensitive attributes that may be revealed as a result of eye tracking discussed here are non-exhaustive---other attributes that may be revealed include age, gender, and race~\cite{liebling2014privacy}. As shown above, existing research indicates that we need to be cautious and better protect gaze and eye tracking data given what their analysis may reveal. However, more research is needed to better understand this area.

\section{Applications and ethics}
Gaze tracking has a number of practical uses in MR. Firstly, it allows for more efficient computation. By knowing where in the scene the user is looking, the computer can render less of the scene at a time, focusing on where the user is looking and the surrounding areas. Efficiency is particularly important in immersive experiences as lag and jitters can worsen nausea and cybersickness~\cite{laviola2000discussion}. Gaze tracking also has the potential to improve accessibility. For example, gaze-based navigation can provide access to users with mobility impairments, allowing them to navigate web pages using their eyes instead of touch-centric interfaces~\cite{hosfelt}.

\subsection{Usability and physical world data}
Usability is an important research focus in the HCI community. One method of usability testing uses eye tracking capabilities to build a heat map of a website, noting how users interact with features.

This type of testing takes time, and recent trends in product development are towards integrating continuous user testing. Imagine a user visiting a web page while using an immersive MR HMD with eye tracking: the web page activates the device's eye tracking capabilities, allowing the site to rapidly iterate with new features, allowing for continuous testing. However, this could also violate user privacy expectations by not informing them of their role in the continuous experimentation, especially experimentation that can reveal sensitive information as discussed above.

In this scenario, the main ethical dilemma is the unchecked testing on users who are not able to appropriately consent, according to commonly agreed upon research protocols. The integration of gaze tracking into devices that we use to browse the web make this scenario plausible.

However, on the MR immersive web (incorporating AR as well as VR), the physical and virtual worlds may be overlaid together. In this scenario, not only would the web page know where the user is looking in the MR web page, but they might also know where the user is looking in the physical world. In fact, this might be a requirement in order to prevent users from inadvertently coming to physical harm while immersed in an HMD. In this case, the device is tracking in real-time the user's reactions to the physical world around them without informed consent. However, knowing this additional information about the user, known as data exhaust, can be lucrative for the companies that collect it in a surveillance capitalist world~\cite{zuboff2019age}.

\subsection{Psychology and eye tracking}
The habits and patterns we have in the physical world persist in the virtual world. For example, people make eye contact, even in virtual environments, and turn to look at speakers in virtual social gatherings. Likewise, our physical bodies still react to actions that happen to our virtual avatars when immersed. For example, when slapped in immersive VR, our bodies will still have a similar physiological reaction as if we were just physically assaulted~\cite{slater2010first}. Users and devices form a feedback loop, but only one has the ability to continuously record and store the data for later analysis.

MR only works if the system measures body movements because the content responds accordingly; according to Bailenson, ``spending 20 minutes in a VR simulation leaves just under 2 million unique recordings of body language''~\cite{bailenson2018protecting}. Some of these recordings will be of intentional body movements and communications, such as speech, while others are done unconsciously. Mainstream psychological research uses eye tracking to gain insights into individuals' problem solving and reasoning strategies~\cite{poole2006eye}. In fact, a key challenge to gaze as an input for navigation is that our eyes often make involuntary, subconsciously controlled movements~\cite{jacob2003eye}.

\section{Conclusion}
MR generates huge amounts of data, and can capture intimate moments of our lives. Identifying the specific ways in which eye tracking mechanisms in mixed reality impact privacy allows product and platform creators as well as legislators to implement the most appropriate solutions for their mitigation. 

Companies that collect and process gaze and eye tracking data need to commit to privacy policies that outline the use of data for only necessary purposes, for which it may not be sold or used for advertising. While it may be tempting to use such a rich source of data for additional purposes, the potential for revealing sensitive health-related and private information is too high. Beyond this, regulators and commercial entities must work together to understand and protect gaze and other biometrically-derived data from misuse and unethical data sharing practices.

\balance{} 

\bibliographystyle{SIGCHI-Reference-Format}
\bibliography{sample}

\end{document}